\documentclass[aps,prl,preprint,superscriptaddress,showpacs,showkeys]{revtex4-1}

\usepackage{amsmath}
\usepackage{amsfonts}
\usepackage{amssymb}
\usepackage[utf8]{inputenc}
\usepackage[T2A]{fontenc}
\usepackage{bm}

\begin{document}

\title{GENERALIZED SPIN-ORBIT INTERACTION\\  IN
TWO-DIMENSIONAL ELECTRON SYSTEMS}

\author{A. A. Eremko}
\email[]{eremko@bitp.kiev.ua}
\affiliation{Bogolyubov Institute for Theoretical Physics, Nat. Acad. of Sci. of Ukraine,
 Kyiv,  Ukraine}
\noaffiliation
\author{L. Brizhik}
\email[]{brizhik@bitp.kiev.ua}
\altaffiliation{}
\affiliation{Bogolyubov Institute for Theoretical Physics, Nat. Acad. of Sci. of Ukraine, Kyiv,  Ukraine}
\author{V. M. Loktev}
\email[]{vloktev@bitp.kiev.ua}
\affiliation{Bogolyubov Institute for Theoretical Physics, Nat. Acad. of Sci. of Ukraine,
 Kyiv,  Ukraine}
\affiliation{National Technical University of Ukraine "Ihor Sikorsky Kyiv Polytechnic
Institute", 
 Kyiv, Ukraine
}

\begin{abstract}
In frame of Dirac quantum field theory that describes electrons and
positrons as elementary excitations of the spinor field, the
generalized operator of the spin-orbit interaction is obtained using
non-relativistic approximation in the Hamilton operator of the
spinor field taking into account the presence of an  external
potential.\,\,This operator is shown to contain a new term in
addition to the known ones.\,\,By an example of a model potential in
the form of a quantum well, it is demonstrated that the
Schr\"{o}dinger equation with the generalized spin-orbit interaction
operator describes all spin states obtained directly from the Dirac
equation.\,\,The dependence of the spin-orbit interaction on the
spin states in quasi-two-dimensional systems of electrons localized
in a quantum well is analyzed.\,\,It is demonstrated that the
electric current in the quantum well layer induces the spin
polarization of charge carriers near the boundary surfaces of the
layer, with the polarization of the charge carriers being opposite
at the different surfaces.\,\,This phenomenon appears due to the
spin-orbit interaction and is known as the spin Hall effect, which
was observed experimentally in heterostructures with the
corresponding geometry.
\end{abstract}



\maketitle

\section{Introduction}\vspace*{-1mm}

It is generally accepted that both the origin and the magnitude of
the spin-orbit interaction (SOI), which occurs in various electron
systems, are well-known and profoundly studied
\cite{Bethe,Davydov}.\,\,In principle, such a viewpoint does not
raise objections, and one may agree that SOI is of an exclusively
relativistic origin.\,\,The concept of SOI was introduced into the
Schr\"{o}dinger equation (SE) on the basis of some empirical
(actually, classic) considerations, and it is known as the Thomas
correction \cite{Thomas}.\,\, It is worth to recall
that this correction was independently and simultaneously proposed
by Ya.I.~Frenkel \cite{Frenkel} and it would be
proper to call it the Thomas--Fren\-kel correction.

Since the SOI effect
is purely relativistic, its comprehensive description can be obtained
only in framework of Dirac theory.\,\,As a rule, the known
relativistic corrections to the SE (one of them was called the SOI
operator) are obtained either as the non-relativistic limit
transition in the Dirac equation (DE) or using the
Foldy--Wouthuysen approximation in the Dirac Hamiltonian with respect to the small parameter $v/c$,
where $v$ is the characteristic velocity of the particle, and $c$ is the speed
of light. 

In spite of that, the authors of \cite{AoP1, AoP2} obtained the form of the SOI operator on the basis of the
direct solution of the DE, in which the external potential was taken
into account from the very beginning.\,\,Ad\-di\-tio\-nal\-ly, the
attention was paid to the fact that there are several operators~--
these are the so-called spin invariants~-- that commute with the DE,
but do not commute with one another.\,\,From whence, it follows that the sought
solutions of the DE (even those obtained in the framework of
perturbation theory with respect to the same small parameter) can be
different.\,\,In other words, this means that the DE has a number of
solutions, and, hence, there may exist other corrections to the
SE.\,\,Indeed, it turned out that the corresponding physical
solutions, namely the eigenfunctions and eigenvalues, can be determined,
which behave differently under different circumstances,
i.e.\,\,depending on the potential form and shape, the geometry of
the system, and so on.\,\,Ho\-we\-ver, the eigenvalues and
eigenfunctions found in the cited works do not provide an ultimate
solution to the problem, because there arises a problem of
determining the form of corresponding corrections or the SOI
operator in the SE, considering the spin invariants which those
solutions correspond to.

This problem is addressed in  the present paper within the Dirac
quantum field theory that describes electrons and positrons as
elementary excitations of a spinor field.\,\,By applying the
non-relativistic approximation to the Hamilton operator of the
spinor field in the presence of an external potential,
a generalized SOI operator can be found.\,\,By an example of a
quantum-well (QW) model potential, it is shown that the SE with the
generalized SOI operator describes all spin  states obtained
directly from the DE itself \cite{AoP1,AoP2}.\,\,The dependence of
SOI on the spin state in quasi-two-dimensional (2D) electron systems
localized in the QW plane is analyzed.\,\,Recall that the
consideration of SOI and the study of its effects in low-dimensional
systems (for example, various heterostructures) is a challenging
problem of modern theoretical and applied solid-state physics.

\section{Hamiltonian of Particles in External Fields}

In quantum field theory, the Hamiltonian of the Dirac spinor field has the
form of the integral \cite{QuantField,RelQuant}
\begin{equation}
\mathrm{H}=\int\mathcal{H}d\mathbf{r}=\int\Psi^{\dagger}(\mathbf{r},t)\hat
{H}_{D}\Psi(\mathbf{r},t)d\mathbf{r}, \label{H_SF}%
\end{equation}
where the integration is carried out over the entire volume, and the
four-component function $\Psi(\mathbf{r},t)=$ $=\left(
\psi_{1}:\psi_{2}:\psi _{3}:\psi_{4}\right)  ^{T}$ (the 4-spinor or
bispinor) is the amplitude of the spinor field.\,\,The function
$\Psi(\mathbf{r},t)$ satisfes the DE\vspace*{-1mm}
\begin{equation}
i\hbar\frac{\partial\Psi}{\partial t}=\hat{H}_{D}\Psi\label{DE}%
\end{equation}
with the Hamiltonian\vspace*{-1mm}
\begin{equation}
\hat{H}_{D}=c\left(\!  \hat{\mathbf{p}}-\frac{e}{c}\mathbf{A}(\mathbf{r}%
)\!\right)
\bm{\hat{\alpha}}+e\Phi(\mathbf{r})\hat{I}+mc^{2}\hat{\beta}.
\label{H_D}%
\end{equation}
Here, $\hat{\mathbf{p}}=-i\hbar\bm{\nabla}$ is the momentum operator for a
particle with the mass $m$ and the charge $e$; $\bm{\hat{\alpha}}=\sum
_{j}\mathbf{e}_{j}\hat{\alpha}_{j}$ is a vector matrix, whose components
$\hat{\alpha}_{j}$ ($j=x,y,z$) together with the matrix $\hat{\beta}$ and the
unit matrix $\hat{I}$ are Hermitian Dirac 4-matrices; and $\mathbf{A}%
(\mathbf{r})$ and $\Phi(\mathbf{r})$ are the vector and scalar,
respectively, potentials of the external electromagnetic
field.\,\,Ac\-cor\-ding to quantum mechanics, the components of the
function $\Psi$ are $q$-numbers, i.e.
$\psi_{\nu}^{\dagger}\psi_{\mu}\neq\psi_{\mu}\psi_{\nu}^{\dagger}$,
where $\left(  \nu,\mu\right)  =1,2,3,4$, and the operator
$\hat{H}_{D}$ in Eqs.~(\ref{H_SF}) and (\ref{H_D}) is called the
Dirac Hamiltonian.

Since an arbitrary bispinor $\Psi(\mathbf{r})$ can be expanded in a
complete basis of orthonormal bispinors
$\Psi_{\{\nu\}}(\mathbf{r})$,\vspace*{-1mm}
\[
\Psi(\mathbf{r})=\sum_{\{\nu\}}\hat{c}_{\{\nu\}}\Psi_{\{\nu\}}(\mathbf{r}),
\]
then Hamiltonian (\ref{H_SF}) can easily be written in terms of the
creation, $\hat{c}_{\{\nu\}}^{\dagger}$ and annihilation,
$\hat{c}_{\{\nu\}}$, operators, which are characterized by a set of
quantum numbers $\{\nu\}$.\,\,The natural choice for this basis is
the eigenbispinors of the equation\vspace*{-1mm}
\begin{equation}
\hat{H}_{D}^{(0)}\Psi^{(0)}(\mathbf{r})=E\Psi^{(0)}(\mathbf{r}%
),\label{mH_D(0)}%
\end{equation}
with the Hamilton operator \vspace*{-1mm}
\[
\hat{H}_{D}^{(0)}=c\hat{\mathbf{p}}\bm{\hat{\alpha}}+mc^{2}\hat{\beta},
\]
which describe stationary states of free particles in the absence of
external fields.\,\,In this case, it is convenient to use plane
waves\vspace*{-1mm}
\begin{equation}
\Psi(\mathbf{r})=\Psi\left(  \mathbf{k}\right)  \exp\left(  i\mathbf{k}%
\mathbf{r}\right)  \label{plainwaiv}%
\end{equation}
as the eigenfunctions of the operator $\hat{\mathbf{p}}$, which is
an integral of motion for problem (\ref{mH_D(0)}).\,\,The components
of the wave vector
$\mathbf{k}$ determine the eigenvalues of the momentum $\mathbf{p}%
=\hbar\mathbf{k}$ and belong to the set of quantum numbers
$\{\nu\}$.

If the orthonormalization conditions are imposed, it is easier to
work with a discrete spectrum, rather than with a continuous
one.\,\,With this aim in view, let us confine the space by a cube
with the edge length $L$.\,\,Then, by imposing the cyclic boundary
conditions on the solutions, we get the quasicontinuous
spectrum of the vector $\mathbf{k}=\sum_{j}%
k_{j}\mathbf{e}_{j}$ components:\vspace*{-1mm}
\[
k_{j}=\frac{2\pi}{L}n_{j},\quad n_{j}=0,\pm1,\pm2, ...,\quad
j=x,y,z.
\]

Substituting Eq.~(\ref{plainwaiv}) into Eq.~(\ref{mH_D(0)}),
we obtain the matrix equation\vspace*{-1mm}
\begin{equation}
\label{H_D_m} \left(\!\! \begin{array}{cc} m c^{2} \hat{I}_{2} &
\hbar c \mathbf{k}\bm{\hat{\sigma}} \\ \hbar c
\mathbf{k}\bm{\hat{\sigma}} & -m c^{2} \hat{I}_{2}
\end{array} \!\!\right)\!\! \left( \!\!\begin{array}{c}
\psi_{u} \\
\psi_{d}
\end{array}\!\! \right) = E \left(\!\! \begin{array}{c}
\psi_{u} \\
\psi_{d}
\end{array} \!\!\right)\!\! ,
\end{equation}
where $\bm{\hat{\sigma}}$ is a vector matrix, whose components are
the Pauli matrices, and $\hat{I}_{2}$ is a unit matrix of the second
rank.\,\,In this case, the eigenbispinor of Eq.~(\ref{mH_D(0)})
looks like $\Psi^{(0)}\left( \mathbf{k}\right)  =\left(
\psi_{u}\left( \mathbf{k}\right)  \,\psi _{d}\left(
\mathbf{k}\right)  \right) ^{T}$, where $\psi_{u}=$ \mbox{$=\left(
\psi_{1}\,\psi_{2}\right)  ^{T}$} and $\psi_{d}=\left(
\psi_{3}\,\psi _{4}\right)  ^{T}$ are its upper and lower,
respectively, spinors.

It is rather easy to find a solution of Eq.~(\ref{H_D_m}), the
orthonormalized eigenbispinors of which are given by the expressions
\cite{Bethe,Davydov}\vspace*{-1mm}
\begin{equation}
\label{solutn}
\begin{array}{ll}
\displaystyle \Psi^{(0)}_{e,\sigma}\left( \mathbf{k}\right) =
A_{\mathbf{k}} \left(\!\! \begin{array}{c} \chi_{e,\sigma} \\[1mm]
\dfrac{\hbar c\mathbf{k}\cdot \bm{\hat{\sigma}}}{\varepsilon
\left(\mathbf{k} \right)
 + mc^{2}} \chi_{e,\sigma}
\end{array} \!\!\right)\!\! , &  E = \varepsilon \left(\mathbf{k} \right)\! , \\[8mm]
\displaystyle \Psi^{(0)}_{p,\sigma}\left( \mathbf{k}\right) =
A_{\mathbf{k}} \left(\!\! \begin{array}{c} -\dfrac{\hbar
c\mathbf{k}\cdot \bm{\hat{\sigma}}}{\varepsilon \left(\mathbf{k}
\right)
 + mc^{2}} \chi_{p,\sigma} \\ \chi_{p,\sigma}
\end{array} \!\! \right)\!\! , & E = - \varepsilon \left(\mathbf{k} \right)\! ,
\end{array}\!\!\!\!\!\!\!\!\!\!
\end{equation}
in which the following notations were introduced:\vspace*{-1mm}
\begin{equation}
\label{eps(k)} A_{\mathbf{k}} = \sqrt{\frac{\varepsilon
\left(\mathbf{k} \right) + mc^{2}}{2 \varepsilon \left(\mathbf{k}
\right)}} ,\quad \varepsilon \left(\mathbf{k} \right) =
\sqrt{m^{2}c^{4} + c^{2}\hbar^{2}. \mathbf{k}^{2}}.\!\!\!\!
\end{equation}
The bispinors $\Psi_{e,\sigma}^{(0)}$ and $\Psi_{p,\sigma}^{(0)}$ in
Eq.~(\ref{solutn}) are the amplitudes of the fields of particles
(electrons, $e$) and antiparticles (positrons, $p$), respectively.

Equation (\ref{H_D_m}) has four eigenvectors, because it is the
equation for the eigenvalues of a 4-matrix.\,\,The\-re\-fore, the
number $\sigma$ was introduced in solutions (\ref{solutn}).\,\,This
parameter has two values that are attributed to a pair of orthogonal
spinors $\chi_{\nu,\sigma}$'s: $\chi_{\nu,\sigma
}^{\dagger}\chi_{\nu,\sigma^{\prime}}=\delta_{\sigma,\sigma^{\prime}}$.\,\,Thus,
system (\ref{solutn}) determines a complete set of bispinors, which
can be used in order to represent an arbitrary bispinor in the
form\vspace*{-1mm}
\begin{equation}
\label{finPhi} \Psi \left( \mathbf{r} \right) = \frac{1}{L^{3/2}}
\sum_{\mathbf{k} ,\sigma} e^{i\mathbf{k} \mathbf{r}}
 \left(\! a_{\mathbf{k},\sigma} \Psi^{(0)}_{e,\sigma}\left( \mathbf{k}\right) + b^{\dagger}_{-\mathbf{k},\sigma}
\Psi^{(0)}_{p,\sigma}\left( \mathbf{k}\right) \!\right)\!\!,
\end{equation}\vspace*{-5mm}

\noindent
where $a_{\mathbf{k},\sigma}(b_{\mathbf{k},\sigma})$ and $a_{\mathbf{k}%
,\sigma}^{\dagger}(b_{\mathbf{k},\sigma}^{\dagger})$ are the
operators of the particle (antiparticle) creation and annihilation,
respectively.\,\,The physical condition of positive definiteness
imposed on the eigenvalues of Hamiltonian (\ref{H_SF}) requires that
those operators have to obey the Fermi statistics.

Nevertheless, the definition of bispinors in Eq.~(\ref{finPhi}) is
ambiguous, because bispinors (\ref{solutn}) satisfy
Eq.~(\ref{H_D_m}) for \textit{arbitrary} spinors $\chi_{\nu,\sigma}$
($\nu=e,p$).\,\,The\-re\-fore, the physical meaning of $\sigma$ as a
quantum number is absent.\,\,Al\-though the values of the quantity
$\sigma$ can be defined as $\pm1$ (or $\uparrow$, $\downarrow$),
their meaning as projections on that or another axis still remains
absent, because the directions of those axis are not given, although
it is required for  the complete determination of the state.

At the same time, the stationary states of the system are known to
be characterized by quantum numbers that correspond to a complete
set of observed quantities and have definite values.\,\,The
operators of those quantities, which are called invariants, commute
both with the Hamiltonian of the system and with one another.\,\,As
was said above, for expressions (\ref{solutn}), such numbers are the
eigenvalues of the operator $\hat{\mathbf{p}}$ or the components of
the vector $\mathbf{k}$.

Concerning the spin number, there are several spin invariants in a
uniform space \cite{Sokolov}, which together with the Hamiltonian
have a common system of eigenfunctions (\ref{solutn}).\,\,The
substitution of expressions (\ref{solutn}) into the equation for the
invariant eigenvalues brings about the equations\vspace*{-1mm}
\begin{equation}
\label{egnspinor_p,a} \mathbf{u}_{\nu} \left( \mathbf{k} \right)
\bm{\hat{\sigma}} \chi_{\nu ,\mathbf{k},\sigma} = \sigma u_{\nu}
\left( \mathbf{k} \right) \chi_{\nu ,\mathbf{k},\sigma} ,  \,
u_{\nu}\left( \mathbf{k} \right) = | \mathbf{u}_{\nu}\left(
\mathbf{k} \right)| ,
\end{equation}\vspace*{-5mm}

\noindent which determine both a pair of orthogonal spinors and a
coordinate system, in which they have the simplest form,
$\chi_{\uparrow }=(  1~0)  ^{T}$ and $\chi_{\downarrow}=( 0~1)
^{T}$.\,\,The subscripts $\nu$ and $\sigma$ in
Eq.~(\ref{egnspinor_p,a}) take the relevant values: $\nu=e$ or $p$,
and $\sigma=\pm1$.\,\,Then the matrices $\mathbf{u}_{\nu}(
\mathbf{k}) \bm{\hat{\sigma}}$ become independent invariants
 for
particles ($\nu=e$) and antiparticles ($\nu =p$)\, (worth recalling that in
the general case the vectors $\mathbf{u}_{e}( \mathbf{k})  $ and
$\mathbf{u}_{p}( \mathbf{k})  $ do not coincide), so that
Eq.~(\ref{egnspinor_p,a}) gives sense to the number $\sigma=\pm1$ in
the operators $a_{\mathbf{k},\sigma}^{\dagger}$ and
$a_{\mathbf{k},\sigma}$ in Eq.~(\ref{finPhi}) by indicating that
their action leads to the creation or annihilation of a particle
with a \textit{definite} spin invariant value at a given
$\mathbf{u}_{\nu}(  \mathbf{k})  $.\,\,Eve\-ry invariant has its
own vectors $\mathbf{u}_{\nu}(  \mathbf{k}%
)  $ with a characteristic dependence on \textbf{k}, the account of
which is mandatory.\,\,Note that the spin operator itself is not an
integral of motion even in a uniform space.\,\,Therefore, in the
case of states with a given energy, one may talk about the spin only
as about the mean value of the corresponding operator.

Since the spin invariants do not commute with one another, only one
of them can correspond to the stationary spin state.\,\,The presence
of several invariants gives rise to an ambiguous choice of the spin
state, which is an actual origin of the spinor arbitrariness in
Eq.~(\ref{solutn}).

Important is the fact that the spin state is not defined \textit{a
priori}.\,\,In the general case, an arbitrary linear combination of
all invariants, whose coefficients are free parameters, can be
chosen as such an invariant.\,\,The explicit expressions for the
vectors $\mathbf{u}_{e}\left( \mathbf{k}\right)  $ and
$\mathbf{u}_{p}\left(  \mathbf{k}\right)  $ (k) are given in work
\cite{BEL}, where it was shown that they really contain free
parameters or the components of the vectors $\mathbf{u}_{\bm{\mu}}%
\equiv\mathbf{u}_{\bm{\mu}}\left(  \mathbf{k}\right)  $ and $\mathbf{u}%
_{\bm{\mathcal{S}}}\equiv\mathbf{u}_{\bm{\mathcal{S}}}\left(  \mathbf{k}%
\right)  $.\,\,The subscripts $\bm{\mu}$ and $\bm{\mathcal{S}}$ are
associated with the notation for the initial invariants: $\bm{\mu}$
for the vector of \textit{magnetic spin polarization} and
$\bm{\mathcal{S}}$ for the vector of \textit{spin polarization}
\cite{Sokolov}, the linear combination of which
$\hat{\mathcal{I}}_{\rm gen}=\mathbf{r}_{\bm{\mu}}\left(
\mathbf{k}\right) \bm{\hat{\mu}}+\mathbf{r}_{\bm{\mathcal{S}}}\left(
\mathbf{k}\right) \bm{\hat{\mathcal{S}}}$ is an invariant in the
general form.\,\,In the coordinate space,
$\mathbf{k}\rightarrow(1/\hbar)\hat{\mathbf{p}}$, and the vectors
$\mathbf{u}_{\bm{\mu}}$ and $\mathbf{u}_{\bm{\mathcal{S}}}$ become
operators that commute with the Hamiltonian and with the invariants
of the spatial
motion.\,\,The choice of $\mathbf{u}_{\bm{\mu}}$ and $\mathbf{u}%
_{\bm{\mathcal{S}}}$ that explicitly depend on $\hat{\mathbf{p}}$ corresponds
to that or another invariant.\,\,For example, if $\mathbf{u}_{\bm{\mu}}%
=\hat{\mathbf{p}}$ and $\mathbf{u}_{\bm{\mathcal{S}}}=0$, we deal with the
operator of helicity, whereas if $\mathbf{u}_{\bm{\mu}}=0$ and $\mathbf{u}%
_{\bm{\mathcal{S}}}=\mathbf{e}_{j}\times\hat{\mathbf{p}}$, this is
the corresponding component of the vector of \textit{electric spin
polarization}.\vspace*{-2mm}

\section{Spin-Orbit Interaction}

Expressions (\ref{solutn}), in which the spinors are given by
Eq.~(\ref{egnspinor_p,a}) with the vectors $\mathbf{u}_{e}\left(
\mathbf{k}\right)  $ and $\mathbf{u}_{p}\left(  \mathbf{k}\right)  $, give an
explicit form of the bispinors $\Psi_{e,\sigma}^{(0)}\left(  \mathbf{k}%
\right)  $ in expansion (\ref{finPhi}).\,\,The substitution of
Eq.~(\ref{finPhi}) into Eq.~(\ref{H_SF}) leads to the Hamiltonian of
the Dirac spinor field in
the representation of free-particle occupation numbers,%
\[
\mathrm{H} = \sum_{\mathbf{k},\sigma} \biggl(\!
\varepsilon(\mathbf{k}) a_{\mathbf{k},\sigma}^{\dagger}
 a_{\mathbf{k},\sigma}\, +\, 
 \frac{1}{L^{3}} \sum_{\mathbf{k}',\sigma'} V_{\sigma ,\sigma'}^{(e-e)} \left(\mathbf{k},
  \mathbf{k}' \right) a_{\mathbf{k},\sigma}^{\dagger} a_{\mathbf{k}',\sigma'}
 \!\! \biggr) +
+ \sum_{\mathbf{k},\sigma} \biggl(\! - \varepsilon(\mathbf{k})
b_{-\mathbf{k},\sigma} b^{\dagger}_{-\mathbf{k},\sigma}\,
 +
\]\vspace*{-3mm}
\[
+\, \frac{1}{L^{3}} \sum_{\mathbf{q},\sigma'} V_{\sigma
,\sigma'}^{(p-p)}\left(\mathbf{k}, \mathbf{k}'
 \right) b_{-\mathbf{k},\sigma} b^{\dagger}_{-\mathbf{k}',\sigma'}\! \biggr)  +
+\, \frac{1}{L^{3}} \sum_{\mathbf{k}, \mathbf{k}',\sigma ,\sigma'}
\biggl(\! V_{\sigma ,\sigma'}^{(e-p)} \left(\mathbf{k}, \mathbf{k}'
\right) a^{\dagger}_{\mathbf{k},\sigma}
b^{\dagger}_{-\mathbf{k}',\sigma'}\, +
\]\vspace*{-3mm}
\begin{equation}
\label{H_rel1}
 +\,  V_{\sigma ,\sigma'}^{(p-e)} \left(\mathbf{k}, \mathbf{k}' \right) b_{-\mathbf{k},\sigma}
 a_{\mathbf{k}',\sigma'} \!\!\biggr)\!,
\end{equation}
where
\begin{equation}
\label{calV_p,a} 
 V_{\sigma ,\sigma'}^{(\nu-\nu')} \left(\mathbf{k},
\mathbf{k}' \right) = \left(\! \Psi^{(0)}_{\nu,\sigma} \left(
\mathbf{k}\right) \!\right)^{\!\dagger} \times
\left(\!  V \left(\mathbf{k} - \mathbf{k}'
\right) \hat{I} - \frac{e}{c}
 \mathbf{A} (\mathbf{k} - \mathbf{k}') \bm{\hat{\alpha}} \!\right) \Psi^{(0)}_{\nu',\sigma'}\left( \mathbf{k}'\right)
\end{equation}
are the particle-particle ($\nu=\nu^{\prime}=e$),
antiparticle-antiparticle ($\nu=\nu^{\prime}=p$), and
particle-antiparticle ($\nu\neq\nu^{\prime}$) scattering matrices,
which contain the Fourier images of the scalar, $V\left(
\mathbf{r}\right)  $, and vector, $\mathbf{A}(\mathbf{r})$,
potentials, i.e.\,\,$V\left(  \mathbf{q}\right)  $ and
$\mathbf{A}\left(  \mathbf{q}\right)  $,
respectively.\,\,Then, operator (\ref{H_rel1}) becomes the sum $\mathrm{H}%
=\mathrm{H}_{e}+\mathrm{H}_{p}+\mathrm{V}_{e-p}$ of three terms: the
particle, $\mathrm{H}_{e}$, and antiparticle, $\mathrm{H}_{p},$
Hamiltonians, and the operator $\mathrm{V}_{e-p}$ of direct
particle-antiparticle transformation. If the product of the creation
and annihilation operators is in the normal form, Hamiltonian
(\ref{H_rel1}) becomes positively determined, except for an infinite
additive constant, which is the energy of the state without any
particles (\textquotedblleft vacuum\textquotedblright).\,\,This is
the energy, from which  the energies of all elementary excitations
are reckoned from.

The state of the system with the given number of elementary excitations is described by
the ket vector $|\psi\rangle$ that satisfies the SE%
\[
i\hbar \frac{\partial}{\partial t} \vert \psi \left( t \right)
\rangle = \mathrm{H} \vert \psi \left( t \right)\rangle.
\]
Here, in accordance with the considered problem and the conservation
laws, the state $|\psi\rangle$ is generated by the products of the
required numbers of the particle and antiparticle creation operators
that act on the vacuum state $|0\rangle$.\,\,In a uniform isotropic
space, Hamiltonian (\ref{H_rel1}) has a diagonal form and is the sum
of the Hamiltonians of free particles and antiparticles.\,\,In
addition, it also separates the Hamiltonians for particles with
opposite spins.\,\,As one can see, in the presence of external
fields, (i)~the independence of the free particle and antiparticle
states disappears, and (ii)~the elements of the scattering matrix
(\ref{calV_p,a}) are not only generated by the scalar and vector
potentials, but also depend on the form of spinors
$\chi_{\nu,\mathbf{k},\sigma}$ in amplitudes (\ref{solutn}).\,\,The
explicit dependence of spinors (\ref{egnspinor_p,a}) on the wave
vector points directly at the interrelation between the spin and
spatial levels of the particle degrees of freedom.\,\,For the
description of this interrelation, the SOI concept was
introduced.\,\,We should emphasize that, hence, SOI is nothing else,
but a direct result of the presence of that or another external
potential violating the uniform rectilinear motion of particles.

\vspace*{-0.5mm}Here we confine the consideration to the case
when the magnetic field is absent, by putting
$\mathbf{A}(\mathbf{r})=0$ in Eq.~(\ref{calV_p,a}).\,\,As was shown
in work \cite{BEL}, in the case of non-relativistic potentials,
where the inequality \mbox{$|V( \mathbf{r}) |/mc^{2}\ll1$} takes place, the particle and antiparticle states can be approximately
separated with a given accuracy making use of the canonical
Schrieffer--Wolf transformation method.\,\,In the majority of
physical problems, the kinetic energy is also a non-relativistic
quantity, so that another inequality, \mbox{$\hbar k/mc\ll1$}, is
obeyed, which allows energies (\ref{eps(k)}) and bispinor
convolutions in Hamiltonian (\ref{H_rel1}) to be expanded in series
in this small parameter.\,\,The approximate renormalization with
respect to those both parameters makes it possible to change to the
non-relativistic approximation in Hamiltonian
(\ref{H_rel1}) and represent it as the sum $\mathrm{H}=\tilde{\mathrm{H}}%
_{e}+\tilde{\mathrm{H}}_{p}$, where the components $\tilde{\mathrm{H}}%
_{\nu=e,p}$ describe already independent (quasi)particles and
(quasi)antiparticles, which include a small (within the second
order of magnitude) \textquotedblleft admixture\textquotedblright\ of the
initial states of both particles and antiparticles.

\vspace*{-0.5mm}Below, only the electron Hamiltonian is considered,
so that the subscript \textquotedblleft$e$\textquotedblright\ is
omitted.\,\,The analysis of the antiparticle case is almost
identical to the presented one.\,\,In particular, in work
\cite{BEL}, both parameters were assumed to be of the same order of
magnitude, and a non-relativistic approximation for Hamiltonian
(\ref{H_rel1}) was obtained within the second-order
corrections.\,\,Thus, it was shown that the non-relativistic
Hamiltonian $\tilde{\mathrm{H}}_{e}$ for
electrons in an external scalar potential looks like%
\begin{equation}
\label{H_nrl-1}
\tilde{\mathrm{H}}_{e} \equiv \mathrm{H} \simeq
\sum_{\mathbf{k},\sigma} \biggl( \! \frac{\hbar^{2}
\mathbf{k}^{2}}{2m}
 \left( 1 - \lambda_{\rm SO} k^{2} \right) a_{\mathbf{k},\sigma}^{\dagger}
 a_{\mathbf{k},\sigma}\,
  + \, \frac{1}{L^{3}} \sum_{\mathbf{k}',\sigma'} V_{\sigma ,\sigma'} \left(\mathbf{k} ,\mathbf{k}' \right)
   a_{\mathbf{k},\sigma}^{\dagger} a_{\mathbf{k}',\sigma'}\!\! \biggr)\!,
\end{equation}
where
\begin{equation}
\label{calV_nr}
V_{\sigma ,\sigma'} \left(\mathbf{k} ,\mathbf{k}' \right) = V
\left(\mathbf{k} - \mathbf{k}' \right)
 \chi_{\sigma}^{\dagger}\,\times
 \left(\! 1 - \frac{\lambda_{\rm SO} }{2} \left( \mathbf{k} - \mathbf{k}' \right)^{2}
  + i \lambda_{\rm SO} \bm{\Lambda} \left( \mathbf{k},\mathbf{k}' \right) \bm{\hat{\sigma}}\! \right) \chi_{\sigma'}\! ,
\end{equation}
and the parameter
\begin{equation}
\lambda_{\rm SO}=\frac{\hbar^{2}}{4m^{2}c^{2}}\label{l_SO}%
\end{equation}
characterizes the order of relativistic corrections and determines the SOI magnitude.

Hamiltonian (\ref{H_nrl-1}) includes all relativistic corrections of
the second order to both the kinetic and potential energies.\,\,The
second term in the renormalized scattering potential (\ref{calV_nr})
is known as the Darwin correction, and the matrix
$\bm{\Lambda}\left(  \mathbf{k},\mathbf{k}^{\prime }\right)
\bm{\hat{\sigma}}$ is a relativistic correction that is called the
SOI operator.\,\,In the SOI operator, the vector $\bm{\Lambda}\left(
\mathbf{k},\mathbf{k}^{\prime}\right)  $ is described by the expression%
\begin{equation}
 \label{L_tot}
\bm{\Lambda} \left( \mathbf{k},\mathbf{k}' \right) =  \mathbf{k}
\times \mathbf{k}' + \bm{\Lambda}_{\rm BEL}
 \left( \mathbf{k},\mathbf{k}' \right)\! .
 \end{equation}
Here, the term $\mathbf{k}\times\mathbf{k}^{\prime}$ corresponds to
the Thomas--Frenkel correction (see, e.g., 
\cite{Bethe,Davydov,RelQuant}).\,\,The other term,
$\bm{\Lambda}_{\rm BEL}\left(  \mathbf{k},\mathbf{k}^{\prime
}\right) $, is given by the expression
\begin{equation}
\bm{\Lambda}_{\rm BEL}\left(  \mathbf{k},\mathbf{k}^{\prime}\right)
=\bm{\Lambda}\left(  \mathbf{k}\right)  -\bm{\Lambda}\left(  \mathbf{k}%
^{\prime}\right)\!  ,\label{L_BEL}%
\end{equation}
where the vector\vspace*{-3mm}
\begin{equation}
\bm{\Lambda}\left(  \mathbf{k}\right)  =\mathbf{e}\times\mathbf{u}%
^{(2)}\left(  \mathbf{k}\right)  +\frac{\mathbf{e}\cdot\left[  \mathbf{e}%
_{z}\times\mathbf{u}^{(2)}\left(  \mathbf{k}\right)  \right]  }{1+\mathbf{e}%
\mathbf{e}_{z}}\mathbf{e}\label{L(k)}%
\end{equation}
was also obtained in work \cite{BEL} in the case where $\mathbf{u}_{e}\left(
\mathbf{k}\right)  \equiv\mathbf{u}\left(  \mathbf{k}\right)  $ has the form
\begin{equation}
\mathbf{u}\left(  \mathbf{k}\right)  \simeq\mathbf{u}^{(0)}+2\lambda
_{\rm SO}\mathbf{u}^{(2)}\left(  \mathbf{k}\right)  +...\,.\label{vecr_nr}%
\end{equation}
In this expansion, $\mathbf{u}^{(0)}$ and $\mathbf{u}^{(2)}$ denote the terms
of the zeroth and second orders, respectively:
\[
\mathbf{u}^{(0)}=\mathbf{u}_{\bm{\mu}}+\mathbf{u}_{\bm{\mathcal{S}}},
\qquad
\mathbf{u}^{(2)}\left(  \mathbf{k}\right)  =\left(  \mathbf{u}%
_{\bm{\mathcal{S}}}\cdot\mathbf{k}\right)  \mathbf{k}-\left[  \mathbf{u}%
_{\bm{\mu}}\times\mathbf{k}\right]  \times\mathbf{k}.
\]
Taking this into account, the solution of Eq.~(\ref{egnspinor_p,a}) can be
written as follows:%
\[
\chi_{\mathbf{k},\sigma}\simeq\chi_{\sigma}+\lambda_{\rm SO}\chi_{\mathbf{k}%
,\sigma}^{(2)}=\left(  1-i\lambda_{\rm SO}\bm{\Lambda}\left(
\mathbf{k}\right) \bm{\hat{\sigma}}\right)  \chi_{\sigma},
\]
where $\bm{\Lambda}\left(  \mathbf{k}\right)  $ is given by expression
(\ref{L(k)}), and $\chi_{\sigma}$ satisfies the spinor equation
\begin{equation}
\mathbf{u}^{(0)}\bm{\hat{\sigma}}\chi_{\sigma}=\sigma u^{(0)}\chi_{\sigma
},\quad\sigma=\pm1.\label{spinor_e}%
\end{equation}
The guiding cosines $\gamma_{j}$ of the vector $\mathbf{u}^{(0)}$
with
respect to the axes of the selected coordinate frame ($\sum_{j}\gamma_{j}%
^{2}=1$) are spin variables.\,\,For example, it is easy to see that
the spinors
\begin{equation}
\label{spinor_tet,fi} \chi_{\sigma} \equiv \chi_{\sigma} \left(
\theta , \phi \right)  = e^{i\sigma \phi /2} \left(
\!\!\begin{array}{c}
\sigma \sqrt{\dfrac{1+\sigma \gamma_{z}}{2}} e^{-i\phi /2} \\[3mm]
\sqrt{\dfrac{1-\sigma \gamma_{z}}{2}} e^{i\phi /2}
\end{array}\!\! \right)\!\!,
\end{equation}
where $
\tan\phi=\frac{\gamma_{y}}{\gamma_{x}},\quad\gamma_{z}=\cos\theta,
$
are the solutions of Eq.~(\ref{spinor_e}) in the laboratory coordinate frame
($j=x,y,z$).\,\,In this case, the unit vector $\mathbf{e}=\mathbf{u}%
^{(0)}/u^{(0)}$ in expression (\ref{L(k)}) can be written in the parametric
form (in terms of spin variables, which are the arguments of spinors
(\ref{spinor_tet,fi})) as follows:
\begin{equation}
\mathbf{e}(\mathbf{k}_{\perp})=\sin\theta\cos\phi\mathbf{e}_{x}+\sin\theta
\sin\phi\mathbf{e}_{y}+\cos\theta\mathbf{e}_{z}.\label{e_qspin}%
\end{equation}

In view of expression (\ref{L(k)}), the vector $\bm{\Lambda}\left(
\mathbf{k},\mathbf{k}^{\prime}\right),$ which characterizes SOI in the
non-relativistic Hamiltonian, can be written in the form
\begin{equation}
\bm{\Lambda}\left(  \mathbf{k},\mathbf{k}^{\prime}\right)  =\mathbf{k}%
\times\mathbf{k}^{\prime}+\mathbf{e}\times\bm{\lambda}\left(  \mathbf{k}%
,\mathbf{k}^{\prime}\right)  +\frac{\mathbf{e}\cdot\left[  \mathbf{e}%
_{z}\times\bm{\lambda}\left(  \mathbf{k},\mathbf{k}^{\prime}\right)  \right]
}{1+\mathbf{e}\mathbf{e}_{z}}\mathbf{e}, \label{Lambda_k,k'}%
\end{equation}
where $\bm{\lambda}\left(  \mathbf{k},\mathbf{k}^{\prime}\right)
=\mathbf{u}^{(2)}\left(  \mathbf{k}\right)  -\mathbf{u}^{(2)}\left(
\mathbf{k}^{\prime}\right)  $, and the vector $\mathbf{u}^{(2)}\left(
\mathbf{k}\right)  $ is defined in Eq.~(\ref{vecr_nr}).

Note that the operators $a_{\mathbf{k},\sigma}^{\dagger}$ and $a_{\mathbf{k}%
^{\prime},\sigma}$ in Eq.~(\ref{H_nrl-1}) are related to the spinors
$\chi_{\sigma}$, which satisfy equality (\ref{spinor_e}), and
describe the creation and annihilation, respectively, of electrons
with the spin polarization determined by the vector
$\mathbf{u}^{(0)}$ [see Eq.~(\ref{vecr_nr})].\,\,As a result, the
vector \textbf{e} that enters expression (\ref{Lambda_k,k'}) inserts
not only the dependence on spatial variables into the SOI operator,
but also the explicit dependence on the spin degrees of
freedom.\,\,In the external potential, the stationary electron
states will be realized only provided a known spin invariant.\,\,In
the non-relativistic approximation, this is the matrix
$\mathbf{u}^{(0)}\bm{\hat{\sigma}}$.\,\,Ac\-cor\-ding to the
symmetry of the given potential, this spin invariant is
related to the vectors $\mathbf{u}_{\bm{\mu}}$ and $\mathbf{u}%
_{\bm{\mathcal{S}}}$ with a dependence on \textbf{k} (in the coordinate
representation, on the momentum), which is to be specified.

\section{Diagonalization of the Hamiltonian Regarding Spin-Orbit Interaction}

It is well known that the relativistic effects in the problems of
non-relativistic physics can be taken into account by adding only
the relativistic SOI operator, which describes spin-dependent
phenomena, to the standard Schr\"{o}dinger Hamiltonian.\,\,Bea\-ring
all that in mind and taking Eqs.~(\ref{calV_nr}) and
(\ref{Lambda_k,k'}) into account, let us write down Hamiltonian
(\ref{H_nrl-1}) in the form\vspace*{-1mm}
\[
\mathrm{H} = \sum_{\mathbf{k},\sigma} \biggl( \frac{\hbar^{2}
\mathbf{k}^{2}}{2m} a_{\mathbf{k},\sigma}^{\dagger}
 a_{\mathbf{k},\sigma}\, +
\, \frac{1}{L^{3}} \sum_{\mathbf{k}'} V \left(\mathbf{k} - \mathbf{k}' \right)
 a_{\mathbf{k},\sigma}^{\dagger} a_{\mathbf{k}',\sigma}\, +\biggr.
 \]
\begin{equation}
\label{H_nrl_e}
\biggl. +\, i \frac{ 1}{L^{3}}\lambda_{\rm SO} \sum_{\mathbf{k}',\sigma'} V \left(\mathbf{k} - \mathbf{k}' 
 \right) \times\,
 \chi_{\sigma}^{\dagger} \bm{\Lambda} \left( \mathbf{k},\mathbf{k}' \right) \bm{\hat{\sigma}} \chi_{\sigma'}
 a_{\mathbf{k},\sigma}^{\dagger} a_{\mathbf{k}',\sigma'} \biggr) .
\end{equation}

As concerning the eigenstates \mbox{$|\psi\left(  t\right)
\rangle=$} \mbox{$=\exp\left( -i\mathcal{E}t/\hbar\right)
|\psi\rangle$} of electrons in a given potential,
the solution procedure of the SE $\mathrm{H}|\psi\rangle=\mathcal{E}%
|\psi\rangle$ is reduced to the diagonalization of the Hamiltonian
with the help of a unitary transformation\vspace*{-1mm}
\begin{equation}
a_{\mathbf{k},\sigma}=\sum_{\{n\}}\psi_{\{n\},\sigma}\left(  \mathbf{k}%
\right)  a_{\{n\},\sigma}, \label{transf_1}%
\end{equation}
whose coefficients $\psi_{\{n\},\sigma}\left(  \mathbf{k}\right)  $ are
determined by an equation for eigenvalues and satisfy the orthonormalization
condition%
\[
\sum_{\mathbf{k}}\psi_{\{n\},\sigma}^{\ast}\left(  \mathbf{k}\right)
\psi_{\{n^{\prime}\},\sigma}\left(  \mathbf{k}\right)  =\delta
_{\{n\},\{n^{\prime}\}}.
\]
These coefficients are given by a set $\{n\}$ of all quantum numbers that
determine the energy $E_{\{n\},\sigma}$, i.e. by the equation
\[
\frac{\hbar^{2} \mathbf{k}^{2}}{2m} \psi_{\sigma} (\mathbf{k}) +
\frac{1}{L^{3}} \sum_{\mathbf{k}'} V \left( \mathbf{k} - \mathbf{k}'
\right) \psi_{\sigma} (\mathbf{k}')\, +
\]
\begin{equation}
\label{SEq_k} 
+\, i \lambda_{\rm SO} \frac{1}{L^{3}} \sum_{\mathbf{k}',\sigma'} V
\left(\mathbf{k} - \mathbf{k}' \right) \chi_{\sigma}^{\dagger}
\bm{\Lambda} \left( \mathbf{k},\mathbf{k}' \right)\times \bm{\hat{\sigma}} \chi_{\sigma'}
\psi_{\sigma'} (\mathbf{k}') = \mathcal{E} \psi_{\sigma}
(\mathbf{k}) .
\end{equation}

While solving Eq.~(\ref{SEq_k}), it is convenient to change to the coordinate
representation,
\begin{equation}
\psi_{\sigma}(\mathbf{r})=\frac{1}{L^{3/2}}\sum_{\mathbf{k}}e^{i\mathbf{k}%
\mathbf{r}}\psi_{\sigma}(\mathbf{k}). \label{psi_s-r}%
\end{equation}
Then, the substitution $\mathbf{k}\rightarrow\mathbf{k}^{\prime}+\mathbf{q}$
has to be made in the terms that contain $V\left(  \mathbf{k}-\mathbf{k}%
^{\prime}\right)  $ and the double sum over $\mathbf{k}$ and $\mathbf{k}%
^{\prime}$.\,\,As a result, we obtain the equality (see Eq.~(\ref{Lambda_k,k'}))%
\[
\bm{\lambda} \left( \mathbf{k}, \mathbf{k}' \right) \equiv
\bm{\lambda} \left( \mathbf{q}, \mathbf{k}' \right) = \left(
\mathbf{u}_{\bm{\mathcal{S}}} \cdot \mathbf{q} \right) \mathbf{k}'\,
+ \left( \mathbf{u}_{\bm{\mathcal{S}}} \cdot \mathbf{k}' \right)
 \mathbf{q}\left[ \mathbf{u}_{\bm{\mu}} \times \mathbf{q} \right]\times \mathbf{k}'
 - \left[ \mathbf{u}_{\bm{\mu}} \times
  \mathbf{k}' \right] \times \mathbf{q} + \mathbf{u}^{(2)} \left( \mathbf{q} \right)\! ,
\]
where $\mathbf{u}^{(2)}\left(  \mathbf{q}\right)  $ is defined in
Eq.~(\ref{vecr_nr}).\,\,Fi\-nal\-ly, Eq.~(\ref{SEq_k}) rewritten in
the coordinate representation is transformed into a system of SEs
for the spin states, 
\begin{equation}
\label{SEq_r}
\left(\! \frac{ \hat{\mathbf{p}}^{2}}{2m} + V \left( \mathbf{r}
\right)\! \right) \psi_{\sigma} (\mathbf{r})
 + i \lambda_{\rm SO}\, \times
\sum_{\sigma'} \chi_{\sigma}^{\dagger} \bm{\Lambda} \left( \hat{\mathbf{p}},\bm{\nabla}
 V\left( \mathbf{r} \right) \right) \bm{\hat{\sigma}} \chi_{\sigma'} \psi_{\sigma'} (\mathbf{r}) = \mathcal{E}
  \psi_{\sigma} (\mathbf{r}) ,
\end{equation}
where\vspace*{-2mm}
\begin{equation}
 \bm{\Lambda} \left( \hat{\mathbf{p}},\bm{\nabla}
V\left( \mathbf{r} \right) \right) = -\frac{i}{\hbar} \bm{\nabla}
V\left( \mathbf{r} \right) \times \hat{\mathbf{p}} + \mathbf{e}\,
\times\, \bm{\lambda} \left( \hat{\mathbf{p}},\bm{\nabla} V\left(
\mathbf{r} \right) \right) - \frac{ \left[ \mathbf{e} \times
\bm{\lambda} \left( \hat{\mathbf{p}},\bm{\nabla} V\left( \mathbf{r}
\right) \right) \right] \mathbf{e}_{z}}{1 + \mathbf{e}
\mathbf{e}_{z}} \mathbf{e} .\label{L(p)}
\end{equation}
In the last expression, the following notations were
introduced:\vspace*{-2mm}
\begin{equation}
\label{lambda(p)}
\begin{array}{l}
\displaystyle \bm{\lambda} \left( \hat{\mathbf{p}},\bm{\nabla}
V\left( \mathbf{r} \right) \right) = -\frac{i}{\hbar}
\hat{\bm{\lambda}} - \bm{\lambda} \left( \bm{\nabla} V\left(
\mathbf{r} \right) \right)\! ,
\\[2mm]
\displaystyle \hat{\bm{\lambda}} = \left(
\mathbf{u}_{\bm{\mathcal{S}}} \cdot \bm{\nabla} V\left( \mathbf{r}
\right) \right)
 \hat{\mathbf{p}} + \bm{\nabla} V\left( \mathbf{r} \right) \left( \mathbf{u}_{\bm{\mathcal{S}}}
 \cdot \hat{\mathbf{p}} \right) +
+ \left[ \bm{\nabla} V\left( \mathbf{r} \right)
\times \mathbf{u}_{\bm{\mu}} \right]
 \times \hat{\mathbf{p}} +  \bm{\nabla} V\left( \mathbf{r} \right) \times \left[ \mathbf{u}_{\bm{\mu}} \times
 \hat{\mathbf{p}} \right]\! , \\[1mm]
\displaystyle \bm{\lambda} \left( \bm{\nabla} V\left( \mathbf{r}
\right) \right) = \bm{\nabla} \left( \mathbf{u}_{\bm{\mathcal{S}}}
\cdot \bm{\nabla} V\left( \mathbf{r} \right) \right)
 +\, \bm{\nabla} \times \left[ \mathbf{u}_{\bm{\mu}} \times \bm{\nabla} V\left( \mathbf{r} \right) \right]\!.
\end{array}
\end{equation}

Transformation (\ref{transf_1}) with the coefficients determined by
Eq.~(\ref{SEq_r}) brings Hamiltonian (\ref{H_nrl_e}) to the diagonal
form with respect to the spatial degrees of freedom, and the
diagonalization with respect to the spin number means the
determination of spinors $\chi_{\sigma }\left(  \theta,\phi\right)
$ satisfying the condition $\chi_{\sigma
}^{\dagger}\bm{\Lambda}\left(  \hat{\mathbf{p}}\right)
\bm{\hat{\sigma}}\chi
_{\sigma^{\prime}}\sim\delta_{\sigma,\sigma^{\prime}}$.\,\,This
relationship
holds true, if the spinors $\chi_{\sigma}$, which are eigenfunctions of the matrix $\mathbf{e}%
\bm{\hat{\sigma}}$ according to definition (\ref{spinor_e}), are
also eigenfunctions of the matrix $\bm{\Lambda}\left(
\hat{\mathbf{p}}\right)  \bm{\hat{\sigma}}$.\,\,This is
possible, if these mutually independent matrices commute, namely, if%
\begin{equation}
\label{cond_g}
\left[ \mathbf{e} \bm{\hat{\sigma}} , \bm{\Lambda}
\left( \hat{\mathbf{p}} \right) \bm{\hat{\sigma}} \right]  =
\mathbf{e} \bm{\Lambda} \left( \hat{\mathbf{p}} \right) -
\bm{\Lambda} \left( \hat{\mathbf{p}} \right) \mathbf{e}\, 
 +\,  i \left( \mathbf{e} \times \bm{\Lambda} \left(
\hat{\mathbf{p}} \right) - \bm{\Lambda} \left( \hat{\mathbf{p}}
\right) \times \mathbf{e} \right) \bm{\hat{\sigma}} = 0 .
\end{equation}
Equality (\ref{cond_g}) evidently demands that the vectors
$\mathbf{e}$ and $\bm{\Lambda}\left(  \hat{\mathbf{p}}\right)  $
should commute with each other.\,\,Accor\-ding to
Eq.~(\ref{lambda(p)}), the vector $\bm{\lambda}\left(
\hat{\mathbf{p}}\right)  $ has the term $\bm{\lambda}\left(
\bm{\nabla}V\left(  \mathbf{r}\right)  \right)  $, which depends on
coordinates.\,\,Hence, the same term also enters into  the SOI vector
(\ref{L(p)}).\,\,The\-re\-fore, the commutation condition for the
vector operator $\bm{\Lambda}\left(  \hat{\mathbf{p}}\right)  $ and
the unit vector $\mathbf{e}$ (generally speaking, this vector in the
coordinate representation is also an operator that contains
$\hat{\mathbf{p}}$), is valid provided that $\bm{\lambda}\left(
\hat{\mathbf{p}}\right)  $ does not include a vector that would
depend on the spatial coordinates.\,\,This requirement leads to
the condition%
\[
\bm{\lambda} \left( \bm{\nabla} V\left( \mathbf{r} \right) \right) =
\bm{\nabla}
 \left( \mathbf{u}_{\bm{\mathcal{S}}} \cdot \bm{\nabla} V\left( \mathbf{r} \right)
 \right)
   +\, \bm{\nabla} \times \left[ \mathbf{u}_{\bm{\mu}} \times \bm{\nabla} V\left( \mathbf{r} \right) \right] = 0 .
\]
This equality imposes restrictions on the vectors
$\mathbf{u}_{\bm{\mu}}$ and
$\mathbf{u}_{\bm{\mathcal{S}}}$, and it can be satisfied, only if
\begin{equation}
\label{cond_1} \mathbf{r}_{\bm{\mathcal{S}}} \cdot \bm{\nabla}
V\left( \mathbf{r} \right) = 0 , \quad \mathbf{r}_{\bm{\mu}} \times
\bm{\nabla} V\left( \mathbf{r} \right) = 0 ,
\end{equation}
Essencially, these are equations for the vectors
$\mathbf{r}_{\bm{\mu}}$ and $\mathbf{r}_{\bm{\mathcal{S}}}$, and
their solutions depend on the specific field symmetry, which reveals
itself in the field gradient $\bm{\nabla}V\left(  \mathbf{r}\right)
$.

Since the gradient vector is directed along the normal to the equipotential
surface at any of its points, the equality $\bm{\nabla}V\left(  \mathbf{r}%
\right)  =|\bm{\nabla}V\left(  \mathbf{r}\right)  |\mathbf{n}$ holds
true, where $\mathbf{n}$ is a unit normal vector.\,\,Then, at every
point $M$ in the space, the orthogonal basis $\left\{
\mathbf{e}_{1}\left(  \mathit{M}\right)
,\mathbf{e}_{2}\left(  \mathit{M}\right)  ,\mathbf{n}\left(  \mathit{M}%
\right)  \right\}  $, where $\mathbf{e}_{1}\left(  \mathit{M}\right)
$ and $\mathbf{e}_{2}\left(  \mathit{M}\right)  $ are two orthogonal
unit vectors that lie in a plane tangent to the equipotential
surface at the point $M$ [all three vectors are coupled by the
relationship $\mathbf{e}_{1}\left( \mathit{M}\right)
\times\mathbf{e}_{2}\left(  \mathit{M}\right) =\mathbf{n}\left(
\mathit{M}\right)  $], can be used.\,\,Ac\-cor\-ding\-ly, an
arbitrary vector \textbf{a} can be expanded in this basis, $\mathbf{a}%
=\alpha_{1}\mathbf{e}_{1}\left(  \mathit{M}\right)  +\alpha_{2}\mathbf{e}%
_{2}\left(  \mathit{M}\right) \, +$ $+\,\alpha_{3}\mathbf{n}\left(
\mathit{M}\right) $, which is nothing else but an expression for
this vector in the curvilinear coordinate frame that is related to
the potential $V\left(  \mathbf{r}\right) $.\,\,Ac\-cor\-ding to
condition (\ref{cond_1}), we obtain
\begin{equation}
\label{r_S,m} \mathbf{u}_{\bm{\mathcal{S}}} = \alpha_{1}
\mathbf{e}_{1}\left( \mathit{M} \right) + \alpha_{2}
 \mathbf{e}_{2}\left( \mathit{M} \right) ,\quad \mathbf{u}_{\bm{\mu}} = \alpha_{3}
  \mathbf{n}\left( \mathit{M} \right)
\end{equation}
in this coordinate frame, so that
\[
\mathbf{e}=\mathbf{u}_{\bm{\mu}}+\mathbf{u}_{\bm{\mathcal{S}}}=\alpha
_{1}\mathbf{e}_{1}\left(  \mathit{M}\right)  +\alpha_{2}\mathbf{e}_{2}\left(
\mathit{M}\right)  +\alpha_{3}\mathbf{n}\left(  \mathit{M}\right)  ,
\]
where $\alpha_{1}^{2}+\alpha_{2}^{2}+\alpha_{3}^{2}=1$.

Because of the imposed condition, expression (\ref{L(p)}) contains the
equality $\bm{\lambda}\left(  \hat{\mathbf{p}},\bm{\nabla}V\right)
=-(i/\hbar)\hat{\bm{\lambda}},$ where, according to Eq.~(\ref{lambda(p)}),%
\[
\hat{\bm{\lambda}} = \bm{\nabla} V \left(
\mathbf{u}_{\bm{\mathcal{S}}} \cdot \hat{\mathbf{p}} \right) +
\bm{\nabla} V \times \left[ \mathbf{u}_{\bm{\mu}} \times
\hat{\mathbf{p}} \right] =
\]\vspace*{-7mm}
\[
= \vert \bm{\nabla} V\left( \mathbf{r} \right) \vert \bigl\lbrace
\mathbf{n}\left( \mathit{M} \right)
 \left( \mathbf{u}_{\bm{\mathcal{S}}} \cdot \hat{\mathbf{p}} \right) 
+\,  \mathbf{u}_{\bm{\mu}} \left( \mathbf{n}\left( \mathit{M}
\right) \cdot \hat{\mathbf{p}} \right)
 - \hat{\mathbf{p}} \left( \mathbf{n}\left( \mathit{M} \right) \cdot \mathbf{u}_{\bm{\mu}} \right)\! \bigr\rbrace .
\]
Now, with the help of definition (\ref{r_S,m}) and the rules of vector
computation, we obtain%
\[
\hat{\bm{\lambda}} = \vert \bm{\nabla} V\left( \mathbf{r} \right)
\vert \, \mathbf{e}
 \times \left( \mathbf{n}\left( \mathit{M} \right) \times \hat{\mathbf{p}} \right) =
  \mathbf{e} \times \left( \bm{\nabla} V\left( \mathbf{r} \right) \times \hat{\mathbf{p}} \right)\! .
\]
In view of this formula and the equality
$\mathbf{e}^{2}=1$, vector
(\ref{L(p)}), which characterizes SOI in Eqs.~(\ref{SEq_r}), reads
\begin{equation}
\label{L_SOI} \bm{\Lambda} \left( \hat{\mathbf{p}},\bm{\nabla} V
\right) = -\frac{i}{\hbar}  \frac{ \left( \mathbf{e}
 + \mathbf{e}_{z} \right)\cdot \bm{\nabla} V\left( \mathbf{r} \right) \times \hat{\mathbf{p}} }{1 + \mathbf{e}
  \mathbf{e}_{z}  }  \mathbf{e} .
\end{equation}

Hence, condition (\ref{cond_g}) is obeyed automatically for the
vectors $\mathbf{u}_{\bm{\mu}}$ and $\mathbf{u}_{\bm{\mathcal{S}}}$
that satisfy relations (\ref{cond_1}), and the matrix element
$\chi_{\sigma}^{\dagger }\bm{\Lambda}\left(  \hat{\mathbf{p}}\right)
\bm{\hat{\sigma}}\chi
_{\sigma^{\prime}}\sim\delta_{\sigma,\sigma^{\prime}}$.\,\,A direct
consequence of all that is the separation of SE (\ref{SEq_r}) into
equations that are independent for every spin state:
\begin{equation}
\label{SEq_r_s}
\left(\!  \frac{ \hat{\mathbf{p}}^{2}}{2m} + V \left( \mathbf{r}
\right) + \sigma \frac{\lambda_{\rm SO}}{\hbar}
 \frac{ \left( \mathbf{e} + \mathbf{e}_{z} \right) \cdot \bm{\nabla} V\left( \mathbf{r} \right) \times
 \hat{\mathbf{p}}}{1 + \mathbf{e}  \mathbf{e}_{z}}\! \right)
 \times\, \psi_{\sigma} (\mathbf{r}) = \mathcal{E}
 \psi_{\sigma} (\mathbf{r}).
\end{equation}
Their solutions $\psi_{\{n\},\sigma}(\mathbf{r})$ together with the
corresponding eigenvalues $E_{\{n\},\sigma}$ determine the constants in
transformation (\ref{transf_1}),%
\[
\psi_{\{n\},\sigma}(\mathbf{k})=\frac{1}{L^{3/2}}\int e^{-i\mathbf{k}%
\mathbf{r}}\psi_{\{n\},\sigma}(\mathbf{r})d\mathbf{r}.
\]
At the same time, Hamiltonian (\ref{H_nrl_e}) becomes completely diagonalized,
\begin{equation}
\mathrm{H}=\sum_{\{n\},\sigma}\mathcal{E}_{\{n\},\sigma}a_{\{n\},\sigma
}^{\dagger}a_{\{n\},\sigma}. \label{H_diag}%
\end{equation}

Taking the explicit expressions of spinors (\ref{spinor_tet,fi}) into account
and introducing the spinor functions%
\[
\sum_{\sigma} \chi_{\sigma} \psi_{\sigma} (\mathbf{r}) = \left(\!\!
\begin{array}{c}
\cos \frac{\theta}{2} \psi_{+} - e^{-i\phi} \sin \frac{\theta}{2}
\psi_{-}
\\[1mm]
e^{i\phi} \sin \frac{\theta}{2} \psi_{+} + \cos \frac{\theta}{2}
\psi_{-}
\end{array} \!\!\right)
= \bm{\psi} (\mathbf{r}) = \left(\!\! \begin{array}{c}
\psi_{\uparrow} (\mathbf{r}) \\
\psi_{\downarrow} (\mathbf{r})
\end{array}\!\! \right)\!\! ,
\]
the system of equations (\ref{SEq_r}) or (\ref{SEq_r_s}) can be written as a
single stationary Pauli equation $\hat{H}_{\rm P}\bm{\psi}(\mathbf{r}%
)=\mathcal{E}\bm{\psi}(\mathbf{r})$ with the Hamiltonian $\hat{H}_{\rm P}%
=H_{0}+\mathrm{V}_{\rm SO}$, where
\begin{equation}
H_{0}=\frac{\hat{\mathbf{p}}^{2}}{2m}+V\left(  \mathbf{r}\right)  , \qquad
\mathrm{V}_{\rm SO}=\frac{\lambda_{\rm SO}}{\hbar}\frac{\left(  \hat{\mathbf{e}%
}+\mathbf{e}_{z}\right)  \cdot\left[  \bm{\nabla}V\left(  \mathbf{r}\right)
\times\hat{\mathbf{p}}\right]  }{1+\hat{\mathbf{e}}\mathbf{e}_{z}}%
\hat{\mathbf{e}}\bm{\hat{\sigma}}.
\label{H_P}%
\end{equation}
In this case, the vector
$\hat{\mathbf{e}}=\hat{\mathbf{u}}^{(0)}/u^{(0)}$, which enters the
SOI operator and, by definition, characterizes a spin invariant, has
to be consistent with the commutation condition for the matrix
$\hat{\mathbf{u}}^{(0)}\bm{\hat{\sigma}}$ and the
Hamiltonian.\,\,From whence, there arises the natural requirement
\begin{equation}
\left[  \hat{\mathbf{u}}^{(0)},H_{0}\right]  =0. \label{condSOI}%
\end{equation}
In other words, the vector operator $\hat{\mathbf{u}}^{(0)}$, which
defines the generalized SOI operator, has to be an invariant (or a
function of invariants) of the spatial motion in the given potential
$V\left( \mathbf{r}\right)  $.\,\,In this case, every specific
potential is connected with its \textquotedblleft
own\textquotedblright\ invariant (in the general case, not a single
one).\,\,The\-re\-fore, condition (\ref{condSOI}) governs both the
explicit expression for $\hat{\mathbf{u}}^{(0)}$ and, as a result,
the form of the generalized SOI operator that contains not only the
Thomas--Frenkel correction in this potential.

For illustration, let us consider below a model potential in the
form of a QW, in which the bound electron states are formed by
electrons captured by the well and moving freely in the well plane
(2D \mbox{electrons}).

\section{Free 2D Electrons}

Let us analyze the electron states in a QW potential that are
described by Eq.~(\ref{SEq_r_s}).\,\,Let us select the $z$-axis
along the potential change direction, $V\left(
\mathbf{r}\right)=V\left(  z\right)  $, and represent the momentum
operator in the form $\hat{\mathbf{p}}=\hat{\mathbf{p}}_{\perp
}+\hat{p}_{z}\mathbf{e}_{z}$, where $\hat{\mathbf{p}}_{\perp}=\hat{p}%
_{x}\mathbf{e}_{x}+\hat{p}_{y}\mathbf{e}_{y}$.\,\,In this case, we
have $\bm{\nabla}V\left(  z\right)  =\mathbf{e}_{z}dV\left(
z\right)  /dz\equiv
V^{\prime}\left(  z\right)  \mathbf{e}_{z}$, and Eq.~(\ref{SEq_r_s}) reads%
\[
\left(\! \frac{ \hat{\mathbf{p}}^{2}}{2m} + V \left( z \right) +
\sigma \frac{\lambda_{\rm SO}}{\hbar}
 V'\left( z \right) \frac{ \hat{\mathbf{e}} \cdot \mathbf{e}_{z} \times \hat{\mathbf{p}}}{1 + \hat{\mathbf{e}}
 \mathbf{e}_{z}} \!\right)\times
\, \psi_{\sigma} (\mathbf{r}) = \mathcal{E} \psi_{\sigma} (\mathbf{r}) .
\]
It is easy to verify that two momentum components, $\hat{p}_{x}$ and
$\hat {p}_{y}$, remain the integrals of motion for the chosen
potential.\,\,Ac\-cor\-ding\-ly, the states of electrons captured by
the QW will be characterized by a definite value of the momentum
$\hat{\mathbf{p}}_{\perp}$ and will be described by the normalized
wave function
\begin{equation}
\label{Psi_2D}
\displaystyle\psi_{\mathbf{k}_{\perp}}(\mathbf{r}) = L^{-1}
e^{i\mathbf{k}_{\perp} \mathbf{r}_{\perp} }
 \varphi_{\mathbf{k}_{\perp}} (z)  , \qquad
\displaystyle\mathbf{r}_{\perp} = x \mathbf{e}_{x} + y
\mathbf{e}_{y} ,\qquad
\displaystyle\mathbf{k}_{\perp} = k_{x}
\mathbf{e}_{x} + k_{y} \mathbf{e}_{y}  ,
\end{equation}
where $\mathbf{r}_{\perp}=x\mathbf{e}_{x}+y\mathbf{e}_{y}$ and $\mathbf{k}%
_{\perp}=k_{x}\mathbf{e}_{x}+k_{y}\mathbf{e}_{y}$.\,\,Ac\-cor\-ding
to condition (\ref{condSOI}), the vector $\hat{\mathbf{e}}$ depends
only on $\hat {\mathbf{p}}_{\perp}$.\,\,Sub\-sti\-tu\-ting
expression (\ref{Psi_2D}) into
Eq.~(\ref{SEq_r_s}), we obtain the stationary one-dimensional SE%
\begin{equation}
\label{1DSEq}
\biggl(\! - \frac{\hbar^{2}}{2m} \frac{d^{2}}{dz^{2}} +
\frac{\hbar^{2} \mathbf{k}_{\perp}^{2} }{2m}
 + V \left( z \right) 
 +\, \sigma \lambda_{\rm SO} \frac{ \mathbf{e}(\mathbf{k}_{\perp}) \cdot \mathbf{e}_{z} \times
  \mathbf{k}_{\perp}}{1 + \mathbf{e}(\mathbf{k}_{\perp})  \mathbf{e}_{z}} V'\left( z \right) \!\biggr)\times \,
   \varphi_{\mathbf{k}_{\perp}, \sigma}\left( z \right) = \mathcal{E} \varphi_{\mathbf{k}_{\perp}, \sigma}
   ( z ) ,
\end{equation}
in which the fourth term in the parentheses on the left-hand side
describes the generalized SOI.\,\,This equation coincides with the
equation obtained in \cite{AoP2,FNT}, in which an analytic
general solution of the DE for the given problem was sought, as well
as with the equation obtained in work \cite{BEL} on the basis of the
non-relativistic Hamiltonian (\ref{H_nrl_e}).\,\,No additional
conditions are imposed on the vector
$\mathbf{e}(\mathbf{k}_{\perp})$, which defines the electron spin
state, so that it can be chosen with an arbitrary dependence on
$\mathbf{k}_{\perp}$.\,\,This circumstance means that the free 2D
electrons still retain a certain spin freedom, i.e.\,\,their states
remain spin-indefinite.

The last term in the parentheses on the left-hand side of
Eq.~(\ref{1DSEq}) characterizes the influence that the QW edges
exert, by means of the SOI mechanism, on the dynamics of electrons,
depending on their spin state.\,\,The vector $\mathbf{k}_{\perp}$
can be written in the form $\mathbf{k}_{\perp
}=$ $=k_{\perp}\mathbf{e}_{{k}_{\perp}}$, where%
\[
\mathbf{e}_{\mathbf{k}_{\perp}}=\frac{\mathbf{k}_{\perp}}{k_{\perp}%
}=\mathbf{e}_{x}\cos\varphi_{\mathbf{k}_{\perp}}+\mathbf{e}_{y}\sin
\varphi_{\mathbf{k}_{\perp}}.
\]
In addition, let us consider expression (\ref{e_qspin}), in which
\[
\tan\varphi_{\mathbf{k}_{\perp}}=\frac{k_{y}}{k_{x}},\quad k_{\perp}%
=\sqrt{k_{x}^{2}+k_{y}^{2}}.
\]
This procedure makes it possible to obtain the following
expression for the coefficient that determines the magnitude of the generalized
SOI in Eq.~(\ref{1DSEq}):
\begin{equation}
\frac{\mathbf{e}(\mathbf{k}_{\perp})\cdot\mathbf{e}_{z}\times\mathbf{k}%
_{\perp}}{1+\mathbf{e}(\mathbf{k}_{\perp})\cdot\mathbf{e}_{z}}=k_{\perp
}f\left(  \mathbf{k}_{\perp}\right)  ,\label{f(k)}%
\end{equation}
where the function\vspace*{-3mm}
\[
f\left(  \mathbf{k}_{\perp}\right)  \equiv f\left(  \theta,\phi,\varphi
_{\mathbf{k}_{\perp}}\right)  =\frac{\sin\theta\sin\left(  \phi-\varphi
_{\mathbf{k}_{\perp}}\right)  }{1+\cos\theta}%
\]
was introduced.\,\,Ge\-ne\-ral\-ly speaking, the angles $\theta$ and
$\phi$ may depend on the direction of the vector
$\mathbf{k}_{\perp}$, i.e.\,\,$\theta =\theta(\mathbf{k}_{\perp})$
and $\phi=\phi(\mathbf{k}_{\perp})$.

When solving Eq.~(\ref{1DSEq}), the attention should be paid to that
Hamiltonian (\ref{H_nrl_e}) is a non-relativistic approximation, in
which the terms of the order of $\lambda_{\rm SO}$ were
retained.\,\,The\-re\-fore, the solutions themselves will be correct
exactly to this accuracy, which allows the term proportional to
$\lambda_{\rm SO}$ to be considered as a perturbation.\,\,The
solutions of the equation
\begin{equation}
\label{1DSEq_0} - \frac{\hbar^{2} }{2m} \frac{d^{2}\psi \left( z
\right)}{dz^{2}} + V\left( z \right)
 \psi \left( z \right) = \mathcal{E} \psi \left( z \right)
\end{equation}
are used as the zeroth approximation.

It is known that if $\psi_{n}\left(  z\right)  $ and $\mathcal{E}_{n}$ are
eigenfunctions and eigenvalues, respectively, of Eq.~(\ref{1DSEq_0}), then the
solutions of Eq.~(\ref{1DSEq}) obtained for the discrete QW spectrum in the
first order of perturbation theory can be easily written in the form
\begin{equation}
\label{psi_n} \varphi_{\mathbf{k}_{\perp}, \sigma}\left( z \right) =
\psi_{n} \left( z \right)
 + \sigma \lambda_{\rm SO} k_{\perp} f \left( \mathbf{k}_{\perp} \right)
  \sum_{n'\neq n} c_{n,n'} \psi_{n'} \left( z \right)\! ,
\end{equation}\vspace*{-7mm}
\begin{equation}
\label{E(k)_2D} \mathcal{E}_{n ,\sigma} \left( \mathbf{k}_{\perp}
\right) = \mathcal{E}_{n} + \frac{\hbar^{2}
 \mathbf{k}_{\perp}^{2} }{2m} + \sigma \lambda_{\rm SO} k_{\perp} v_{nn} f \left( \mathbf{k}_{\perp} \right) .
\end{equation}
Here,\vspace*{-3mm}
\begin{equation}
c_{n,n'} = \frac{ v_{nn'}^{\ast} }{\mathcal{E}_{n} -
\mathcal{E}_{n'}}  ,\label{w(n,n)}
\end{equation}\vspace*{-4mm}
\[
v_{n,n'} = \int \psi_{n}^{\ast} \left( z \right) \frac{dV(z)}{dz}
\psi_{n'} \left( z \right) dz ,~~~~
  v_{n,n'}^{\ast} = v_{n',n} .
\]

Thus, the 2D electrons are described by Hamiltonian (\ref{H_diag}), which takes
the form
\begin{equation}
\mathrm{H}_{2D}=\sum_{n,\mathbf{k}_{\perp},\sigma}\mathcal{E}_{n,\sigma
}\left(  \mathbf{k}_{\perp}\right)  a_{n,\mathbf{k}_{\perp},\sigma}^{\dagger
}a_{n,\mathbf{k}_{\perp},\sigma}.\label{H_2D}%
\end{equation}
In this expression, the operators
$a_{n,\mathbf{k}_{\perp},\sigma}^{\dagger}$
($a_{n,\mathbf{k}_{\perp},\sigma}$) are the creation (annihilation)
operators of \textquotedblleft free\textquotedblright\ 2D
electrons\,  with the wave vector
$\mathbf{k}_{\perp}$ in the 2D $\mathcal{E}_{n,\sigma}\left(
\mathbf{k}_{\perp}\right)  $ band associated with the $n$-th
discrete level in the QW, and the spin state of those electrons is
determined by the vector $\mathbf{e}(\mathbf{k}_{\perp})$ [see
Eq.~(\ref{e_qspin})]. In this case, the word \textquotedblleft
free\textquotedblright\ means that those particles are captured by
the QW potential, but move freely in its plane.

From expressions (\ref{E(k)_2D}) and (\ref{psi_n}), one can see that the
generalized SOI gives rise, firstly, to the spin splitting (the Rashba effect)
of 2D bands and, secondly, to the spin dependence of the electron density
distribution over the QW thickness:%
\[
|\varphi_{\mathbf{k}_{\perp},\sigma}\left(  z\right)  |^{2}=|\psi_{n}%
(z)|^{2}+\sigma2\lambda_{\rm SO}k_{\perp}f\left(
\mathbf{k}_{\perp}\right) F_{n}(z),
\]
where\vspace*{-2mm}%
\[
F_{n}(z)=\sum_{n^{\prime}\neq n}\frac{\psi_{n}\left(  z\right)  w_{n,n^{\prime
}}\psi_{n^{\prime}}\left(  z\right)  }{\mathcal{E}_{n}-\mathcal{E}_{n^{\prime
}}}.
\]
The latter phenomenon has not been mentioned earlier.\,\,The spin
splitting of 2D bands (\ref{E(k)_2D}) by a magnitude of
$2\lambda_{\rm SO}k_{\perp}v_{nn}f\left( \mathbf{k}_{\perp}\right) $
is non-zero, only if the inverse potential symmetry is
violated.\,\,The multiplier $v_{nn}\neq0$ in front of
\mbox{$V_{0}\left( -z\right)  \neq V_{0}\left(  z\right)  $}
characterizes the QW asymmetry (both a probable inherent one and,
e.g., that induced by an external electric field applied
perpendicularly to the $xy$-plane).\,\,The spatial separation of
densities for electrons with different spins is characterized by the
function $2\lambda_{\rm SO}k_{\perp}f\left(
\mathbf{k}_{\perp}\right) F_{n}(z)$, which is finite in the
symmetric QW as well, when the Rashba splitting is absent.\,\,Of
course, this is a commonly known fact, and we report the
corresponding results only as an evidence that the generalized SOI
reproduces the well-known and recognized effects.\,\,In addition, it
allows one to verify, to which of the spin-invariants the potential
should correspond in order that, e.g., the Rashba effect takes
place.

\section{Manifestation of the Spin-Orbit Interaction in a 2D System}

\vspace*{-1.5mm}

It is well known that the bound electron states in a QW are
described by a discrete non-degenerate spectrum of SE 
(\ref{1DSEq_0}) with real-valued eigenfunctions $\psi_{n}(z)$ that are characterized by a definite parity.\,\,The even and odd
functions alternate at that with the growth of their eigenenergies.
If the QW is symmetric with respect to the coordinate origin, the
matrix elements (\ref{w(n,n)}) differ from zero only between the
even and odd states.\,\,The\-re\-fore, $F_{n}(-z)=-F_{n}(z)$, and
the distribution of the probability to find an electron with the
wave vector $\mathbf{k}_{\perp}$ and the spin number $\sigma$ turns
out asymmetric as the electron shifts toward either of the QW
edges.\,\,In this case, it is easy to see that the electrons with
the same $\mathbf{k}_{\perp}$ but opposite spins become shifted to
different surfaces, which directly testifies to the appearance of
the spin Hall effect in this situation and, in essence, explains its
mechanism.

At the same time, the spin splitting of energy bands and the
spin-dependent electron distribution over the QW thickness depend on
the QW form, which governs the explicit form of the solutions of
Eq.~(\ref{1DSEq}) or (\ref{1DSEq_0}).\,\,In this situation, the
function $f\left(  \mathbf{k}_{\perp }\right)  $ plays its role and
introduces an explicit dependence on the spin states of electrons
into such phenomena.\,\,Real\-ly, according to Eq.~(\ref{f(k)} ),
the arguments of this function are the spin (the angles $\theta$ and
$\phi $) and spatial (the angle $\varphi_{\mathbf{k}_{\perp}}$)
variables.\,\,The specific values of the spin variables $\theta$ and
$\phi$ correspond to definite spin invariants, which are preserved
in the field  $V(z)$.\,\,In particular, if the $z$-component of the
electric spin polarization $\epsilon_{z}$ is selected as the spin
invariant, then $\theta=\pi/2$ and
$\phi=\pi/2+\varphi_{\mathbf{k}_{\perp}}$ (state~I).\,\,On the other
hand, if the component of the spin pseudovector $\mathcal{S}_{j}$ in
the $xy$-plane is selected as the spin invariant, then
$\theta=\pi/2$, whereas $\phi =\mathrm{const}$ within the interval
$0\leq\phi\leq\pi/2$ (state~II; the value $\phi=0$ corresponds to
the $x$-component of the invariant, and the value $\phi=\pi/2$ to
the $y$-component).

From definition (\ref{f(k)}), one can see that the influence of the
obtained SOI on the dynamics of 2D electrons takes place, if the
projection of $\mathbf{e}(\mathbf{k}_{\perp})$ onto the direction of
the unit vector
$\mathbf{e}_{z}\times\mathbf{e}_{\mathbf{k}_{\perp}}$ differs from
zero.\,\,This influence reaches its maximum, when
$\mathbf{e}(\mathbf{k}_{\perp})$ coincides with the latter vector
(the Rashba spin state).\,\,The realization of a particular spin
state has to be determined by the given conditions (the
concentration of charge carriers, the presence of electric and/or
magnetic fields, the external pressure, the properties of specific
interface, and so forth) and, therefore, has to manifest itself in
real physical experiments.

As was noted in work \cite{AoP2}, in an isolated 2D band (for
example, when free electrons fill the ground state, $n=0$, QW level;
only this case will be considered below), the total energy of
$N_{e}$ electrons described by the equilibrium distribution function
\begin{equation}
\bar{n}_{\sigma,\mathbf{k}}=\frac{1}{\exp\{(\mathcal{E}_{\sigma,\mathbf{k}%
}-\mu)/\mathit{k}_{B}T\}+1}=\bar{n}\left(  \mathcal{E}_{\sigma,\mathbf{k}%
}\right)  \label{f_0}%
\end{equation}
equals
\[
E_{\rm
tot}=\sum_{\sigma,\mathbf{k}_{\perp}}\mathcal{E}_{\sigma}\left(
\mathbf{k}_{\perp}\right)  \bar{n}_{\sigma\mathbf{k}_{\perp}},\quad
\sum_{\sigma,\mathbf{k}_{\perp}}\bar{n}_{\sigma\mathbf{k}_{\perp}}=N_{e},
\]
and Rashba state~I has the lowest energy in the absence of external
fields and at the low temperature $T$.\,\,The Rashba splitting of 2D
bands and their spin polarization were experimentally confirmed for
a number of materials and structures, in which the charge carriers
possess 2D properties (see, e.g., \mbox{review \cite{Bihl}}).

The spatial spin-separation of charge carriers in the QW can affect
the observed local spin value $\mathbf{S}$ or the related magnetic
moment $\mathbf{M}=(e/mc)\mathbf{S}$ that characterizes the 2D
electron system.\,\,The spin \textit{per se}, $\mathbf{as}$ was
mentioned above, has no definite value for stationary states,
because only its absolute value is the integral of
motion.\,\,The\-re\-fore, the observed spin value is given by the
average value of the spin operator.\,\,In the quantum field theory,
the latter together with
Hamiltonian (\ref{H_SF}) is given by the expression%
\[
\mathbf{S} = \frac{\hbar}{2} \int \Psi^{\dagger}(\mathbf{r})
\bm{\hat{\Sigma}} \Psi (\mathbf{r}) d\mathbf{r} ,\quad
\bm{\hat{\Sigma}} = \left(\!\!
\begin{array}{cc}
 \bm{\hat{\sigma}} & 0 \\ 0 & \bm{\hat{\sigma}}
\end{array}\!\! \right) \!\! .
\]
Here, the expressions for the bispinors are given in
Eq.~(\ref{finPhi}).\,\,Un\-like the Hamiltonian, the operator
$\mathbf{S}$ is not diagonalized in the free-particle
representation, but contains terms that couple the creation and
annihilation operators of electrons and positrons with different
spin numbers.\,\,But, when calculating the average value
$\langle\mathbf{S}\rangle =\mathrm{Sp}\left(
\hat{\rho}_{e}\mathbf{S}\right)  $ with the use of the statistical
operator $\hat{\rho}_{e}$ for a system of electrons described by
Hamiltonian (\ref{H_2D}), the non-diagonal terms vanish, and the
following expression is obtained in the non-relativistic
approximation for the spin density
$\langle\hat{\mathbf{s}}\rangle=\langle\hat{\mathbf{S}}\rangle/L^{2}$
in the QW 
\[
\langle\hat{\mathbf{s}}\rangle=\int\langle\hat{\mathbf{s}}(z)\rangle dz,
\]
where\vspace*{-3mm}%
\[
\langle\hat{\mathbf{s}}(z)\rangle=\frac{\hbar}{2L^{2}}\sum_{\mathbf{k}_{\perp
},\sigma}\chi_{\sigma}^{\dagger}\bm{\hat{\sigma}}\chi_{\sigma}|\varphi
_{\mathbf{k}_{\perp},\sigma}\left(  z\right)  |^{2}\bar{n}_{\sigma
,\mathbf{k}_{\perp}}\!.
\]
Here, $\bar{n}_\sigma,\textbf{k}_\perp=\textrm{Sp}\left(  \hat{\rho}%
_ea_\sigma,\textbf{k}_\perp^\dagger a_\sigma,\textbf{k}_\perp\right)  $ is the
electron distribution function, and the spinors $\chi_\sigma$ are defined by
expressions (\ref{spinor_tet,fi}). To be more precise, in the non-relativistic
approximation, when the quantities proportional to $\lambda_{\rm
SO}$ are preserved, the Pauli matrix has to be
substituted by the matrix
$\bm{\hat{\tilde{\sigma}}}=\bm{\hat{\sigma}}+2\lambda_{\rm SO}\left(
\bm{\Lambda}\left(  \mathbf{k}\right)  \times\bm{\hat{\sigma}}+\mathbf{k}%
\times\left[  \mathbf{k}\times\bm{\hat{\sigma}}\right]  \right)
$.\,\,Then the relativistic corrections along the spin direction,
which is determined by an
essentially non-relativistic expression, are small and can be neglected.

Now, by applying the explicit form of spinors, we obtain%
\[
\chi_{\sigma}^{\dagger} \bm{\hat{\sigma}} \chi_{\sigma}= \sigma (
\sin \theta \cos \phi \mathbf{e}_{x} + \sin \theta \sin \phi
\mathbf{e}_{y}\,
+\, \cos \theta \mathbf{e}_{z} ) = \sigma
\mathbf{e}(\mathbf{k}_{\perp}) .
\]
Passing from the summation over the quasicontinuous $\mathbf{k}%
_{\perp}$ variable to the integration, we obtain the following expression
for the average local magnitude of the spin vector density of electrons in the
isolated 2D band:%
\[
\langle \hat{\mathbf{s}} (z) \rangle =\frac{\hbar }{ 8\pi^{2} } \int
dk_{x} dk_{y} \sum_{\sigma} \sigma \mathbf{e}\left(
\mathbf{k}_{\perp} \right) \vert \varphi_{\sigma
,\mathbf{k}_{\perp}}\left( z \right)
 \vert^{2} \bar{n}_{\sigma ,\mathbf{k}_{\perp}}  . \]

In a symmetric QW ($v_{n,n}=0$), the spin splitting of the bands is
absent, and the distribution function $\bar{n}_{\mathbf{k}_{\perp}}$
does not depend on the spin number.\,\,Ta\-king this fact into
account, after summing over the spin index and substituting the
explicit expression for $|\varphi_{\sigma
,\mathbf{k}_{\perp}}\left(  z\right)  |^{2}$, we obtain%
\[
\langle \hat{\mathbf{s}} (z) \rangle =
\frac{\hbar }{ 2\pi^{2} } \lambda_{\rm SO} \int\limits_{0}^{2\pi}  d
\varphi_{\mathbf{k}_{\perp}} \int\limits_{0}^{\infty} d k_{\perp}
 \mathbf{e}\left( \mathbf{k}_{\perp} \right) k_{\perp}^{2}
  f \left( \mathbf{k}_{\perp} \right) \bar{n}_{\mathbf{k}_{\perp}}  F(z)  .
\]

One can see that the average value
$\langle\hat{\mathbf{s}}(z)\rangle$ depends on the given electron
spin state, which determines the specific values of the angles
$\theta$ and $\phi$ in expressions (\ref{e_qspin}) and
(\ref{f(k)}).\,\,In the equilibrium 2D electron gas characterized by
the distribution function (\ref{f_0}), we have
$\bar{n}_{\mathbf{k}_{\perp}}=\bar{n}\left(
\mathcal{E}_{\mathbf{k}_{\perp}}\right)  $, and the electron spins
become compensated over the whole QW thickness,
$\langle\hat{\mathbf{s}}(z)\rangle =0$, in both indicated states~I
and II.

The situation changes radically, if an external electric field
induces a current in the system.\,\,The external field, whose
strength vector lies in the QW plane, perturbs the electron
subsystem and changes its distribution function.\,\,In the linear
approximation with respect to the perturbation, this function can be
written as the sum \cite{Ziman}\vspace*{-1mm}
\[
\bar{n}_{\mathbf{k}}^{\mathrm{pert}}=\bar{n}\left(  \mathcal{E}_{\mathbf{k}%
}\right)  +\Delta n_{\mathbf{k}},
\]
in which the correction associated with the field,\vspace*{-1mm}
\begin{equation}
\label{g_k2D} \Delta n_{\mathbf{k}} = \frac{\hbar \tau e}{m}
\mathbf{E} \cdot \mathbf{k} \left(\! - \frac{\partial \bar{n}\left(
\mathcal{E}_{\mathbf{k}} \right)} {\partial
\mathcal{E}_{\mathbf{k}}} \!\right)
\end{equation}
contains the phenomenological relaxation time $\tau$.

By selecting the $x$-axis along the field direction, we get
the following expression for the local spin polarization of charge
carriers induced by the joint action of SOI and the electric current
in the QW:\vspace*{-1mm}
\[
\langle \hat{\mathbf{s}} (z) \rangle = \frac{\hbar^{2} \tau e E}{
2\pi^{2} m } \lambda_{\rm SO} \int\limits_{0}^{2\pi}
  d \varphi_{\mathbf{k}_{\perp}}\,\times
\int\limits_{0}^{\infty} d k_{\perp} \mathbf{e}\left( \mathbf{k}_{\perp} \right)
   k_{\perp}^{3} f \left( \mathbf{k}_{\perp} \right) \cos \varphi_{\mathbf{k}_{\perp}}
   \left(\! - \frac{\partial \bar{n}\left( \mathcal{E}_{\mathbf{k}} \right)}
    {\partial \mathcal{E}_{\mathbf{k}}}\! \right)
    F(z).
\]
Both the unit vector $\mathbf{e}\left(  \mathbf{k}_{\perp}\right)  $
and the function $f\left(  \mathbf{k}_{\perp}\right)  $ depend only
on the angle $\varphi_{\mathbf{k}_{\perp}}$.\,\,The\-re\-fore, the
integration over $k_{\perp}$ in this expression can be carried out
by transforming it into the integration over the energy and taking
into account that, at low temperatures, the derivative $\left(
-\partial\bar{n}\left(  \mathcal{E}\right)  /\partial
\mathcal{E}\right)  $ behaves like the $\delta$-function, $\delta
(\mathcal{E}-\mathcal{E}_{\rm F})$, where $\mathcal{E}_{\rm F}$ is
the Fermi energy.\,\,In such a way, we obtain the average
value\vspace*{-1mm}
\[
\langle\hat{\mathbf{s}}(z)\rangle=\frac{\tau eEn_{e}}{\pi}\lambda
_{\rm SO}F(z)\bm{\mathcal{I}},
\]
where $n_{e}$ is the number of electrons per unit QW area,
$F(z)\equiv F_{0}(z)$, and the vector\vspace*{-1mm}
\[
\bm{\mathcal{I}}=\int\limits_{0}^{2\pi}\mathbf{e}\left(
\theta,\phi\right)  f\left(
\theta,\phi,\varphi_{\mathbf{k}_{\perp}}\right)  \cos\varphi_{\mathbf{k}%
_{\perp}}d\varphi_{\mathbf{k}_{\perp}}%
\]
is determined by the spin-invariant, to which the electron states
correspond, i.e.\,\,it depends actually on the specific form of the
general spin invariant $\hat{\mathcal{I}}_{\rm gen}$, which was
introduced above.

If the spin state is given by invariant~I, then the vector $\mathbf{e}%
(\mathbf{k}_{\perp})=-\sin\varphi_{\mathbf{k}_{\perp}}\mathbf{e}_{x}%
+\cos\varphi_{\mathbf{k}_{\perp}}\mathbf{e}_{y}$ and $f\left(  \mathbf{k}%
_{\perp}\right)  =$ $=1$.\,\,From whence, we find that $\bm{\mathcal{I}}=\pi\mathbf{e}_{y}%
$.\,\,For the spin state corresponding to invariant~II, when $\mathbf{e}%
(\mathbf{k}_{\perp})=$
$=\cos\phi\mathbf{e}_{x}+\sin\phi\mathbf{e}_{y}$ and
$f\left(  \mathbf{k}_{\perp}\right)  =\sin\left(  \phi-\varphi_{\mathbf{k}%
_{\perp}}\right)  $, we obtain $\bm{\mathcal{I}}=\pi\left(  \cos
\phi\mathbf{e}_{x}+\sin\phi\mathbf{e}_{y}\right)  \sin\phi$.\,\,It
is evident that the vector $\bm{\mathcal{I}}=0$ at $\phi=0$, and the
same vector $\bm{\mathcal{I}}=\pi\mathbf{e}_{y}$ at $\phi=\pi/2$.

Note that the external electric field directed along the $x$-axis
lowers the system symmetry, and only the $y$-component of the spin
pseudovector remains to be an invariant, $\phi=\pi/2$.\,\,Thus, the
local spin polarization of charge carriers in the QW is described by
the formula\vspace*{-1mm}
\[
\langle\hat{\mathbf{s}}(z)\rangle=\tau eEn_{e}\lambda_{\rm
SO}F(z)\mathbf{e}_{y},
\]
where $F(-z)=-F(z)$ and $F(0)=0$.\,\,Hence, an electric current in
the QW layer induces the spin polarization of charge carriers near
the boundary surfaces of the layer, with this polarization being
opposite at the different surfaces.\,\,This, as was indicated above,
is completely associated with the action of the generalized
SOI.\,\,This phenomenon was called the spin Hall effect, and it was
experimentally observed in structures with a similar geometry
\cite{Kato} (see also review \cite{Sinova}).\,\,In our case, it is a
result of exclusively geometric properties of the examined 2D
structures.\vspace*{-2mm}

\section{Conclusions}

To summarize, in this work on the basis of Dirac  quantum theory of the
spinor field, the generalized SOIspin-orbit interaction operator $\mathrm{V}_{\rm
SO}\sim \bm{\Lambda}( \hat{\mathbf{p}},\bm{\nabla}V)
\bm{\hat{\sigma}}$, given by expressions
(\ref{H_P}) and (\ref{L_SOI}), is obtained.\,\,It is demonstrated that this operator, used in the
non-relativistic SE, provides a consistent description of the
influence of SOI on electrons that move in an external potential $V(
\mathbf{r})  $. \,\,Here the external potential is assumed to be small as compared to the characteristic
electron energy, which is always satisfied at least in the
problems of solid state
physics, as well as in many other physical problems.\,\,In this case, the vec\-tor $\bm{\Lambda}(  \hat{\mathbf{p}%
},\bm{\nabla}V)  $, which, together with the well-known Thomas--Frenkel correction, determines SOI, 
 also contains an
additional contribution (see Eq.~(\ref{L_tot})).

Recall that the Thomas--Frenkel correction appears in vector
(\ref{L_tot}), if the explicit dependence of the lower
(\textquotedblleft small\textquotedblright) spinor on the momentum
is made allowance for.\,\,On the other hand, the additional vector
term $\bm{\Lambda}_{\rm BEL}\left(
\mathbf{k},\mathbf{k}^{\prime}\right) $ emerges owing to an
analogous dependence of the $\Lambda_{\rm SO}$-order corrections to
the upper (\textquotedblleft large\textquotedblright) spinor in the
electron amplitude (bispinor). As to our knowledge, this term was
ignored, as a rule.

It is extremely important that the spin invariants play a
substantial role in finding vector (\ref{L_tot}).\,\,Some of them~--
for example, helicity operator and operator of the vector
of electric spin polarization, as well as
the operator of the total momentum $\mathbf{\hat{J}}=\mathbf{L}+\frac{1}%
{2}\hbar\bm{\hat{\Sigma}}$ and the operator $\hat{K}=(  \hat{\mathbf{L}%
}\bm{\hat{\Sigma}}+\hbar)  \hat{\beta}$ commutating with it, - correspond to the vectors
$\mathbf{u}_{\nu }$ (see Eq.~(\ref{egnspinor_p,a})) without
relativistic corrections. Here $\hat{\mathbf{L}}$ is the operator of angular momentum.

Finally, let us formulate conditions, under which the new, obtained
in this work, correction to the standard and widely used SOI does
not appear.\,\,It occurs, if the potential symmetry preserves one of
the just indicated operators, and the generalized SOI operator
automatically takes the form of the Thomas--Frenkel correction for
the corresponding spin state.\,\,If the potential symmetry does not
violate the preservation of another invariant,
then this spin state (or states) will be lost, if the term $\bm{\Lambda}_{\rm BEL}%
\left(  \mathbf{k},\mathbf{k}^{\prime}\right)  $ in
Eq.~(\ref{L_tot}) is ignored.\,\,This was unambiguously illustrated
by the example of a QW potential, in which the Thomas--Frenkel
correction can describe only the spin state with the Rashba
splitting.\,\,Ho\-we\-ver, we have shown that this correction is not
the only possible one.\,\,We can expect that not
only the Thomas--Frenkel term is responsible for all possible
\textquotedblleft spin\textquotedblright\ consequences of SOI, and
that the generalized SOI will find its application.

\vskip2mm 
{\bf Acknowledgement.} 
\textit{The authors thank Yu.\,\,B.\,Gaididei who recently passed away, 
for discussion of the obtained results and the mechanism of
the spin Hall effect.\,\,We also thank O.I.~Voitenko for technical assistance in paper preparation. The work was carried out in the framework
of the budget program KPKVK~6541230 and the scientific program
0117U00236 of the Department of Physics and Astronomy of the
National Academy of Sciences of Ukraine.}

\end{document}